\documentclass[aps,prapplied,a4paper,10pt,floatfix,twocolumn,showkeys,longbibliography]{revtex4-1}
\usepackage{graphicx}
\usepackage{color}
\usepackage{amsmath}
\usepackage{bm}
\usepackage{array}
\newcolumntype{H}{>{\setbox0=\hbox\bgroup}c<{\egroup}@{}}

\usepackage[pdftex,
colorlinks=true,
a4paper
]{hyperref}

\input{MyMc}

\newcommand{\As}{{\cal A}}
\newcommand{\Idc}{\ensuremath{I_\mathrm{dc}}\xspace}
\newcommand{\Iac}{\ensuremath{I_\mathrm{ac}}\xspace}

\newcommand{\Sec}[1]{\section{#1}}

\let\company\textsc 

\begin{document}

\title{%
  YBa$_2$Cu$_3$O$_7$ Josephson diode operating as a high-efficiency ratchet
}

\author{Christoph Schmid}
\thanks{C.S. and A.J. contributed equally to this work.}
\author{Alireza Jozani}
\thanks{C.S. and A.J. contributed equally to this work.}
\author{Reinhold Kleiner}
\author{Dieter Koelle}
\author{Edward Goldobin}
\email{gold@uni-tuebingen.de}
\affiliation{%
  Physikalisches Institut, Center for Quantum Science (CQ) and LISA$^+$,
  Universit\"at T\"ubingen, Auf der Morgenstelle 14, D-72076 T\"ubingen, Germany
}

\date{%
  \today\ File: \textbf{\jobname.\TeX}
}

\begin{abstract}
Using a focused He$^+$ beam for nanopatterning and writing of Josephson barriers we fabricated specially shaped Josephson junctions of in-line geometry in YBa$_2$Cu$_3$O$_7$ thin film microbridges with an asymmetry ratio of critical currents of opposite polarities (non-reciprocity ratio) $\approx 7$ at optimum magnetic field.
Those \emph{Josephson diodes} were subsequently used as ratchets to rectify an applied ac current into a dc voltage.
We also demonstrate the operation of such a ratchet in the loaded regime, where it produces a nonzero dc output power and yields a thermodynamic efficiency of up to $75\units{\%}$.
The ratchet shows record figures of merit: an output dc voltage of up to $212\units{\mu V}$ and an output power of up to $0.2\units{nW}$.
The device has an essential area $\approx 1\units{\mu m^2}$.
For rectification of quasistatic Gaussian noise, the figures of merit are more modest, however the efficiency can be as high as for the deterministic ac drives within some regimes.
Since the device is based on YBa$_2$Cu$_3$O$_7$, it can operate at temperatures up to $\sim40\units{K}$, where more noise is available for rectification.

\end{abstract}



\maketitle

\Sec{Introduction}
\label{Sec:Intro}

Ratchets, also known as Brownian motors, received a lot of attention two decades ago \cite{Linke02, Reimann02,Haenggi09} stimulated by the investigation of molecular motors in biological systems in the 1990s \cite{Svoboda93,Juelicher97}.
In the simplest model one can imagine a point-like particle moving in one dimension along an asymmetric periodic potential under the action of a deterministic or random applied force (drive) with zero time-average.
The ultimate goal of this device is to rectify the applied force and produce a directed motion of the particle (net transport).
Possible applications range from rectification or mixing of electric signals to the mechanical separation of various (nano-)particles (\eg, viruses) \cite{Linke02,Skaug18}.
A lot of different designs were investigated including asymmetric rocking and flashing potentials, asymmetric and random (noisy) drives with different spectral properties, \etc\cite{Reimann02,Haenggi09}.
However we would like to remind right away that rectification of equilibrium thermal fluctuations (white noise) is forbidden by the second law of thermodynamics \cite{Feynman66}.
Still, it is of basic interest to study how close one can approach this limit and still rectify, and which ingredients in terms of noise parameters, bandwidth, \etc are essential.

Among different realizations of ratchets one of the interesting classes includes Josephson ratchets (JRs).
They have a number of advantages\cite{Beck05,Knufinke12}:
(i) directed motion (of the Josephson phase) results in an average dc voltage $\mean{V}$ (via the second Josephson relation), which is easily detected experimentally;
(ii) Josephson junctions (JJs) are very fast devices, which can operate (capture and rectify deterministic or stochastic drives) in a broad frequency range from dc up to few hundred GHz, capturing a lot of spectral energy in the quasistatic regime;
(iii) by varying the junction design and bath temperature, both overdamped and underdamped regimes are accessible.

In a JR the applied bias current $I$ plays the role of a force acting on the system.
Different realizations of JRs were demonstrated, including superconducting quantum interference device (SQUID) ratchets\cite{deWaele:1969:dcSQUID,Zapata:1996:3JJ-SQUID-Ratchet,Weiss00,Sterck02,Sterck05,Sterck09}, Josephson vortex ratchets based on annular long Josephson junctions (ALJJs)\cite{Carapella01,Beck05,Wang09,Knufinke12} or Josephson junction arrays (JJA)\cite{Falo:1999:JJA-Ratchet,Trias:2000:JJA-Ratchet,Falo:2002:FluxonRatchetPot,Shalom:2005:JJA-Ratchet}, or tunable $\varphi$-JJ ratchets\cite{Menditto16}.
The key parameter that determines the figures of merit in JRs is the asymmetry of the potential. It is defined as the ratio of its maximum slopes (that define depinning forces) for the motion of the particle in the positive and the negative directions. In Josephson junctions this is equal to the ratio of positive and negative critical currents $I_{c+}$ and $I_{c-}$, respectively.
We define the asymmetry $\As$ as a quantity that is positive and larger than 1, \ie,
\begin{equation}
  \As = \begin{cases}
    |I_{c-}|/I_{c+}, &\text{if } I_{c+}<|I_{c-}|,\\
    I_{c+}/|I_{c-}|, &\text{if } I_{c+}>|I_{c-}|.
  \end{cases}
  \label{Eq:As.def}
\end{equation}
%
Intuitively it is clear that the larger the asymmetry $\As$ is, the better the ratchet performs.
A quantitative analysis \cite{Knufinke12,Goldobin:2016:JRatchet.ModelIVC} showed that a large asymmetry allows one to achieve a wide operation range of drive current amplitudes (also known as rectification window), a large counter current (corresponding to a heavy load), against which rectification is still possible, and a large thermodynamic efficiency (ratio of output dc to input ac power).
Thus, to fabricate a practically relevant ratchet one should design a system with high (critical current) asymmetry. An ideal ratchet has infinite asymmetry, \eg, $I_{c+}=0$ (or below the noise level), while $|I_{c-}|$ is finite and well above the noise level, or vice versa. To a first approximation, we aim for $\As \sim 10$. To our knowledge, such JRs were not reported until now with one notable exception \cite{Golod:2022:JosDiode@B=0}.
In addition, previously demonstrated JRs were rather large, see Tab.~\ref{Tab:cmp}, which hampers their integration into micro- or nanoelectronic superconducting circuits.
In terms of their potential future use as nano-rectifiers of fluctuations (noise) and also for connecting many ratchets in series to obtain larger rectified voltages, one would like to down-size a single ratchet to sub-\units[]{\mu m} dimensions.
In addition one would like to have the possibility to operate the JR over a wide range of temperatures $T$.
Obviously, an upper limit in operation temperature is given by the transition temperature $T_c$ of the superconducting material used for fabricating the JRs.

Recently, a new wave of interest emerged in the field of asymmetric (non-reciprocal) superconducting systems, termed ``superconducting diodes''\cite{Ando20,Narita22} or ``Josephson diodes''\cite{Wu22,Jeon:2022:Nb-Pt+YIG-Nb:Diode@B=0, Pal:2022:S-TopoSemiMet-S:Diode,Baumgartner:2022:Al-2DEG-Al:Diode,Paolucci23,Ghosh:2024:HTS-JosDiode,Volkov:2024:TwistModalSC-JosDiode}. However, this term was already mentioned in 1997 in the context of the analysis of fluxon motion in long JJs with a step-like critical current density profile\cite{Krasnov:1997:LJJ.j(x)}.
The superconducting diode is defined as a device with asymmetric critical currents $I_{c+}$ and $I_{c-}$ in positive and negative directions.
In fact, these diodes are nothing else than the ratchets mentioned above. One can also use them as switches, \eg, in digital (logic) circuits or as detectors or mixers, similar to a broad range of applications of semiconducting diodes. Below we will use the word diode to denote a universal device, while the word ratchet will be used as a diode with particular application for rectification of noise or ac signals.
The first advantages of some of the diodes proposed recently is that those are based on specially engineered superconducting materials \cite{Ando20,Wu22,Narita22} and, therefore, can be structured down to the nanoscale, \eg, down to  $50\ldots100\units{nm}$.
Another advantage is that some of them \cite{Wu22,Narita22,Jeon:2022:Nb-Pt+YIG-Nb:Diode@B=0} exhibit an asymmetry even at zero magnetic field.
However, the values of asymmetry that were demonstrated up to now are mostly low. In Tab.~\ref{Tab:cmp} we compare the figures of merit for Josephson diodes. The critical temperature  $T_c$ of the materials used is often below $4\units{K}$ (with one notable exception\cite{Ghosh:2024:HTS-JosDiode}), which prohibits operation even in liquid He at $T=4.2\units{K}$.

The aims of this work are:
(I) to construct a highly asymmetric Josephson diode with a large critical current asymmetry, say, $\As\sim10$;
(II) reduce the essential area of the device to about $1\units{\mu m^2}$ and
(III) implement it using the high-$T_c$ cuprate superconductor YBa$_2$Cu$_3$O$_7$ (YBCO),  with $T_c\approx90\units{K}$ that can operate in a wide temperature range (in our case up to $42\units{K}$).

To implement JJs with $I_c$ asymmetry, we use junctions of in-line geometry, as described in the Ref.~\onlinecite{Barone82}, however in the kinetic inductance limit. Such JJs in an external magnetic field $B$ have a skewed point-symmetric $I_c(B)$ dependence. Thus, there are field values where the critical currents in the positive and negative directions are very different, see Appendix~\ref{Sec:InlineNum} for theoretical background.

\Sec{Fabrication}
\label{Sec:Fabrication}

The fabrication of the ratchet devices starts on a $10\times10 \units{mm^2}$ chip, purchased from \company{Ceraco GmbH}.
The chip consists of a 1-mm-thick (001)-oriented \LSAT (LSAT) substrate onto which a 20-nm-thick CeO$_2$ buffer layer followed by a YBCO film with thickness $d=30\units{nm}$  was epitaxially grown by reactive coevaporation \cite{Kinder97}.
Subsequently a 20-nm-thick Au layer was deposited for electrical contacting.
Micropatterning was done within two lithography steps.
First, we  utilize the MLA100 from \company{Heidelberg Instruments} to pattern 200-$\mu$m-long microbridges with width $W\approx 4 \units{\mu m}$ (connected to larger contact pads for wire bonding) in the maP-1205 photoresist from \company{micro resist technology}.
Using Ar ion beam milling, we etch through all the thin film layers down to the substrate.
Second, we remove the Au layer from the microbridges by means of a wet-etching process using TechniEtch ACI2 from \company{MicroChemicals}.
%

To define the JJs and the circuit geometry we utilized the focused He ion beam (He-FIB) in a \company{Zeiss} Orion NanoFab He ion microscope (HIM) with $30\units{keV}$ He$^+$ ions.
By writing a line across a YBCO bridge with a \emph{moderate} irradiation dose $D\sim 500\ldots700\units{ions/nm}$, the He-FIB irradiation creates a Josephson barrier\cite{Cybart:2015:He-FIB.JJs,Cho18,Mueller19,Karrer:2024:bJJ(t)}, such that the corresponding JJ exhibits\cite{Mueller19} an resistively-shunted junction (RSJ)-like $I$--$V$ characteristic (IVC) with a Stewart-McCumber parameter $\beta_C\sim1$.
The critical current density $j_c$ of such JJs decreases exponentially with increasing dose\cite{Mueller19} $D$, and typically $I_c(T)$ vanishes with increasing $T$ at $T\approx 40\units{K}$ for JJs irradiated with moderate dose\cite{Cybart:2015:He-FIB.JJs,Cho18}.
Instead, by writing a line with a \emph{high} dose $D\sim2000\units{ions/nm}$, one creates a resistive wall (a barrier without supercurrent), which behaves similar to a semiconductor, \ie, its resistance diverges as $T\to0$, reaching a few $\units{M\Omega}$ or above at our main working temperature of $4.2\units{K}$ \cite{Mueller19}.
Transmission electron microscopy analysis shows that on the atomic scale the resistive wall corresponds to damaged (amorphized) YBCO along the irradiated line \cite{Mueller19} and mechanically stressed crystalline YBCO in the $50\ldots100\units{nm}$ vicinity of the amorphized region.\footnote{R.~Hutt, C.~Mag\'en, \etal, unpublished.} Further we use the term amorphous resistive wall (ARW) to remind the reader about both structural and electrical properties.

Before fabricating diodes within any of the prepatterned microbridges of width $W$, we wrote He-FIB lines with different moderate $D$ values across several microbridges to produce a series of 4-$\mu$m-wide test JJs for calibration of the $j_c(D)$ dependence on the chip.

The JJ ratchets were then created  in one He-FIB nanofabrication step to produce JJ barriers and ARWs in the inline JJ design, which is described in the next section.
%
%
The dose for writing the JJ barriers was chosen to obtain the target $j_c$ values (see below).
The ARWs were always written with $D=2000\units{ions/nm}$.

\Sec{Design}
\label{Sec:Design}

%
%
\begin{figure}[!htb]
  \begin{center}
    \includegraphics[width=0.96\columnwidth]{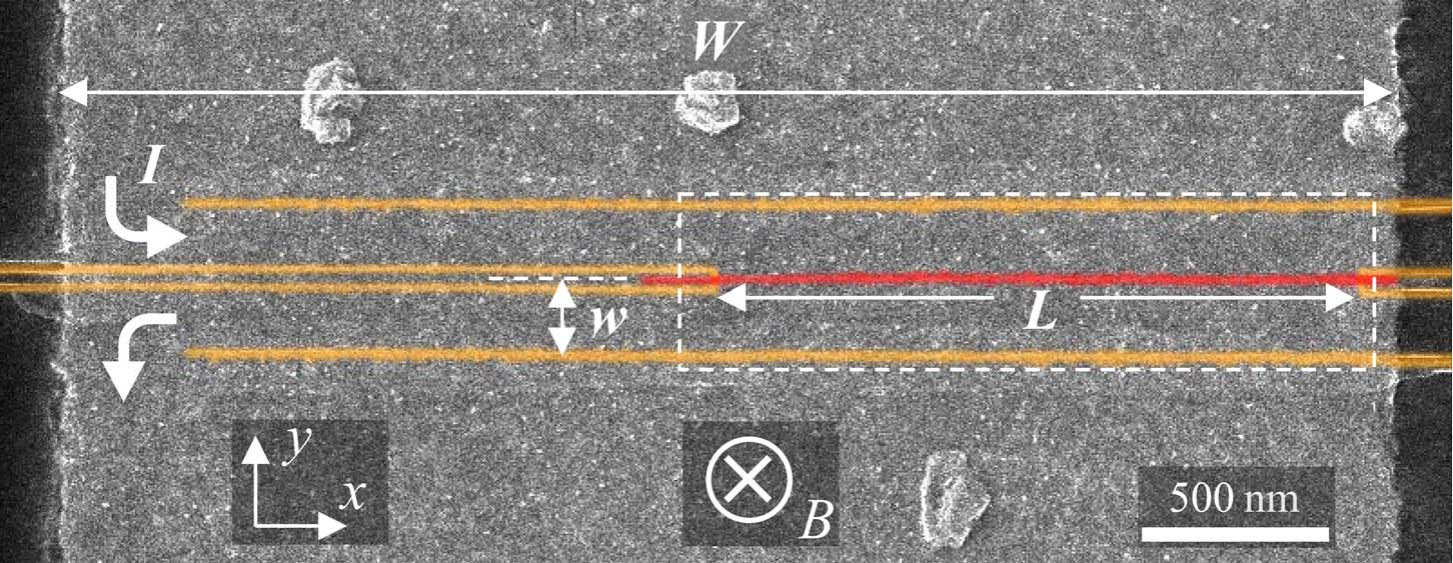}
  \end{center}
  \caption{%
    Secondary electron HIM image of a reference microbridge (\#A15) of the same geometry as device \#A22 discussed in Sec.~\ref{Sec:ExpRes}. For better visibility, \#A15 has been irradiated with a high-dose He-FIB producing an ARW also at the place of the Josephson barrier. The He-FIB lines are colored to indicate where the device \#A22 has ARWs (golden) written by high-dose He-FIB and the Josephson barrier (red) written by moderate-dose He-FIB. The thick arrows show the direction of bias current flow. The width of the whole YBCO bridge is $W$, the JJ length is $L$, and the electrode width (the distance between the JJ barrier and the upper and lower ARWs) is $w$.
    The dashed (white) rectangle shows the essential area $A_\mathrm{e}$ of the diode.
  }
  \label{Fig:Sketch}
\end{figure}

%
One possibility to realize a highly asymmetric ratchet is to use a JJ with an in-line geometry as indicated in the HIM image in Fig.~\ref{Fig:Sketch}. The ARWs compel the bias current flow as indicated by the thick arrows, \ie, the bias current $I$ is injected and extracted from the (left) side, flows parallel to the JJ barrier of length $L$ and across it. The spacing between the barrier and the ARWs is denoted as the electrode width $w$. Usually, for such an in-line geometry one observes a significant self-field effect (non-uniform magnetic field caused by and proportional to the bias current). This results in a skewed dependence of the critical current $I_c$ on the magnetic field $B$ (applied perpendicular to the thin film plane)\cite{Barone82}. In fact, for our very thin YBCO film with thickness $d\ll\lambda_\mathrm{L}$ (typically $\lambda_\mathrm{L}\approx 250\units{nm}$ for YBCO films ), the kinetic inductance dominates. Therefore, it is more correct to speak about a phase gradient (of the macroscopic wavefunctions in the superconducting electrodes) along the barrier (instead of magnetic field) caused by the in-line bias configuration. Similar designs were proposed\cite{Guarcello:2024:inlineLJJ-Diode.T} as diodes very recently and already used for \hstruct{Nb|CuNi|Nb} JJs\cite{Golod:2019:SFS-JJ.SPS} and YBCO grain-boundary JJs \cite{Gerdemann:1995:VFT,Bauch:1997:HTS-GB-JJ:VFT}, but for different purposes.

The planar thin-film JJs used in this work (even in the simplest geometry when the JJ crosses the whole bridge) are non-local \cite{Kogan01}.
Our geometry, see Fig.~\ref{Fig:Sketch}, is even more complicated than the one from Ref.~\onlinecite{Kogan01}.

A theoretical treatment of the $I_c(B)$ dependence for our JJ geometry has not been developed so far. Therefore, the estimations of target parameters for our JJ ratchets are possible only approximately, in the framework of the usual local model, see Appendix \ref{Sec:InlineNum}. There we find that the key parameter that defines the asymmetry of the $I_{c\pm}(B)$ curves is the so-called \emph{in-line geometry parameter}
\begin{equation}
  f_I=\frac{1}{4\pi}\rbfrac{L}{\lambda_J}^2,
  \label{Eq:f_I}
\end{equation}
see Appendix~\ref{Sec:EstLambda_J} for an estimation of $\lambda_J$. There it is shown that
\begin{equation}
  \lambda_J^2 \propto \frac{w}{j_c \lambda_\mathrm{L}^2}
  , \label{Eq:}
\end{equation}
Further, having $f_I$ we can calculate the (skewed) dependences of the normalized critical current $i_{c\pm}=I_{c\pm}/I_{c0}$ ($I_{c0}\equiv j_c L d$) on the applied normalized magnetic flux $f=\Phi/\Phi_0$ ($\Phi_0\approx 2.068\cdot10^{-15}\units{Wb}$ is the magnetic flux quantum) for different values of $f_I$. We find that with increasing $f_I$ the maximum asymmetry $\As_\mathrm{max}$ ($\As$ at an optimum value $f_\mathrm{opt}$ of $f$) increases rapidly  and diverges for $f_I\ge f_I^*=\pi/4$. This happens because the smaller critical current $\min(|I_{c\pm}(B)|)$ vanishes. Experimentally, this divergence is suppressed since there is always a finite background $I_c$ value. Accordingly, we estimate practically achievable values for $\As_\mathrm{max}\sim 10$. For details, see Appendix \ref{Sec:InlineNum}.

Guided by predictions of the local model, we chose the target parameters, such as $D$ (\ie, $j_c$), $L$ and $w$, to achieve the desired asymmetry and reasonable critical current.

The JJ physical length $L$ was chosen to have the in-line geometry parameter $f_I\approx\pi/4$, \ie, $L\approx\pi\lambda_J$ to achieve maximum possible asymmetry $\As\sim10$ at the optimum flux bias point $f_\mathrm{opt}$, see Appendix \ref{Sec:InlineNum}.
In this way JJ also remains in the short JJ limit. Since one of our goals is to make the size of the device as small as possible, we would like to have $L$ as small as possible, \ie, as small as possible $\lambda_J$.

The width $w$ between the ARWs and JJ barrier should be chosen as small as possible to reduce $\lambda_J$ and, accordingly, to reduce the size ($L\times w$) of the diode. However, $w$ is limited by the mechanical damage around the ARWs.

Finally, the dose $D$ was chosen to obtain $j_c$ values that provide JJs with RSJ-like IVC\cite{Mueller19} at $B=0$ and to have a reasonable maximum critical current $I_{c0}=j_c d L$ in the range $5\ldots30\units{\mu A}$ at $T=4.2\units{K}$. Such values of $I_{c0}$ are easily measurable and later allow one to investigate not only the limit of small thermal fluctuations $k_BT/E_J(T) \ll 1$, but also the limit of large thermal fluctuations in a reasonably broad temperature range within out target temperature range of $4.2$--$42\units{K}$. Here $E_J(T)=\Phi_0 I_{c0}(T)/(2\pi)$ is the temperature-dependent Josephson energy. We note that $j_c(T)$ affects $\lambda_J$.

At the end we have fabricated and tested a set of several diodes with the parameters distributed around the target parameters of the local model.

\makeatletter
\newcommand{\ssymbol}[1]{^{\@fnsymbol{#1}}}
\makeatother

\begin{table*}[t]
\centering
\begin{tabular} { l|c c c c c H c c }
  References & Type & ${\cal A}$ & $\mean{V}(\units{\mu V})$ &  $\mean{P}_{\mathrm{out}}(\units{nW})$  &$\eta(\%)$ & size $(\units{\mu m^2})$ & $A_\mathrm{e}(\units{\mu m^2})$ &  $T_{\mathrm{op}}(\units{K})$\\
  \hline
  \hline
  Carapella (2001)\cite{Carapella01} & ALJJ & 1.2 & 5 & - & - && 44500 & 6.5 \\
  Beck (2005)\cite{Beck05} & ALJJ & 2.2 & 20 & - & - && 5700  & 6 \\
  Sterck (2005,2009)\cite{Sterck05,Sterck09} & 3JJ SQUID & 2.5 & 25 & - & - & $45 \times 25$ & $1125$  & 4.2 \\
  Wang (2009)\cite{Wang09} & ALJJ & 2.8 & 100 & - & - && 800  & 4.2 \\
  Knufinke (2012)\cite{Knufinke12,TabNote:Knufi} & ALJJ E3 & 1.6 & 40 & 16$\ssymbol{3}$ & 25\cite{TabNote:EffEst} & $(30+5)^2$ & 4900 & 4.2 \\
  Menditto (2016)\cite{Menditto16} & $\varphi$ junction & 2.5 & 150 & - &- && 2000   & 1.7 \\
  Golod (2022)\cite{Golod:2022:JosDiode@B=0,TabNote:Golod} & in-line JJ & 4 & 8 & - & 70\cite{TabNote:EffEst} & $6 \times 1.2$ & $7.2$ & 7 \\
  Wu (2022)\cite{Wu22} & $\mathrm{NbSe}_2|\mathrm{Nb}_3\mathrm{Br}_8|\mathrm{NbSe}_2$
    & 1.07 & 800\cite{TabNote:RectDrv} & - & 3.4\cite{TabNote:EffEst} & & 3.7 & 0.02\\
  Jeon (2022)\cite{Jeon:2022:Nb-Pt+YIG-Nb:Diode@B=0} & \hstruct{Nb|Pt+YIG|Nb}
    & 2.07 & - & - & 35\cite{TabNote:EffEst} & $\sim2\times2$ & $~4$ & 2\\
  Pal (2022)\cite{Pal:2022:S-TopoSemiMet-S:Diode} & Nb$|$Ti$|$NiTe$_2|$Ti$|$Nb
    & 2.3 & 8\cite{TabNote:RectDrv} & - & 40\cite{TabNote:EffEst} & $\sim 2\times 1.5$ & $\sim3$ & 3.8\\
  Baumgartner (2022)\cite{Baumgartner:2022:Al-2DEG-Al:Diode} & Al$|$2DEG$|$Al
    & 2 & - & - & 30\cite{TabNote:EffEst} & $3.5\times2.1$ & 7 & 0.1\\
  Paolucci (2023)~\cite{Paolucci23} & 2JJ SQUID
    & 3 & 8 &- & 6\cite{TabNote:EffEst} & $6 \times 12$  & $72$ & 0.4 \\
  Gosh (2024)~\cite{Ghosh:2024:HTS-JosDiode,Volkov:2024:TwistModalSC-JosDiode} & twisted BSCCO flakes
    & 4 & 25\cite{TabNote:RectDrv} & - & 60\cite{TabNote:EffEst} & $10\times10$ & $100$ & 80\\
  \hline
  \hline
  This work & in-line JJ & 7 & 212 & 0.2 & 74 & $0.5\times 2$ & 1.0 & 4.2--42 \\
  \hline
\end{tabular}
  \caption{
    Comparison of key parameters of Josephson diodes from literature. $\As$ is the asymmetry \eqref{Eq:As.def}, $\mean{V}$ is the maximum rectified voltage (for optimum drive amplitude $\Iac$), $\mean{P}_{\mathrm{out}}$ is the maximum output power measured, $\eta$ is the maximum thermodynamic efficiency reached.
    $A_\mathrm{e}$ is defined as the area of the part of the device essential for its operation, excluding electrodes, parts that can be safely ``cut off'' without affecting the operation. If devices should be combined into an array, $A_\mathrm{e}$ is the area of one period of such an array. For many cited works we estimated $A_\mathrm{e}$ from the size of the minimum rectangle containing the part of the device essential for its operation.
  }
  \label{Tab:cmp}
\end{table*}

\Sec{Experimental results}
\label{Sec:ExpRes}

Here we present experimental data only for the device \#A22 with $L=1750\units{nm}$ and $w=200\units{nm}$. The barrier was written with $D=530\units{ions/nm}$, which resulted in a maximum $I_{c0}\approx 14\units{\mu A}$ at $T=4.2\units{K}$ (see below). Thus, $j_c\approx 27\units{kA/cm^2}$ and $\lambda_J\approx1.25\units{\mu m}$ at $4.2\units{K}$, see Appendix \ref{Sec:EstLambda_J} for details.

\subsection{Characterization}

\begin{figure}
\includegraphics{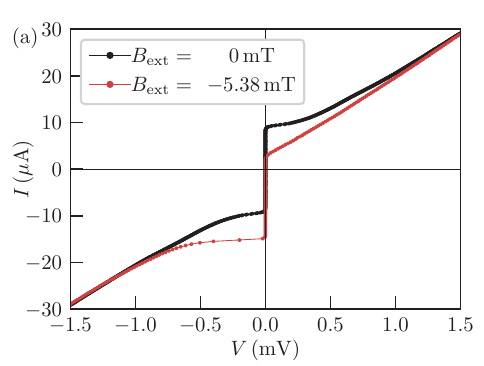}\\
\hspace{5mm}\includegraphics{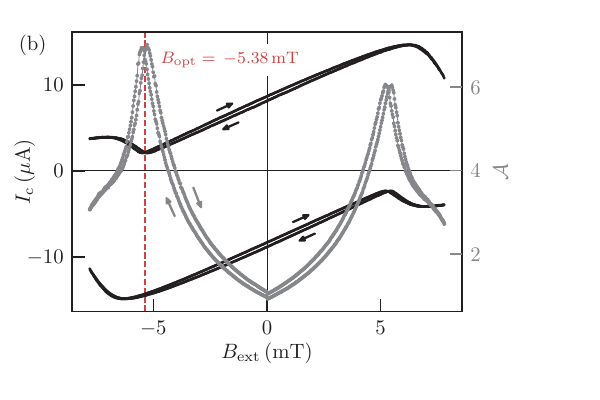}
\caption{%
  Electric transport data for sample \#A22:
  (a) IVCs measured in both  bias-current sweep-directions at $B=0$ (black) and at optimal applied field $B_\mathrm{opt}$ (red).
  $I_\mathrm{c+}(0) \approx |I_\mathrm{c-}(0)| \approx 10\units{\mu A}$, while $I_\mathrm{c+}(B_\mathrm{opt}) \approx 2.0\units{\mu A}$ and $|I_\mathrm{c-}(B_\mathrm{opt})| \approx 14.2\units{\mu A}$ (taken from IVCs with a voltage criterion of $1\units{\mu V}$), which results in $\As(B_\mathrm{opt})\approx 7$.
  Both IVCs are somewhat rounded around the critical currents due to thermal fluctuations and nonuniformities of the JJ, so $I_\mathrm{c\pm}$ could be determined experimentally only approximately.
  (b) $I_\mathrm{c\pm}(B)$ measured in both $B$ sweep directions (black) and $\As(B)$ (grey) numerically calculated from the $I_\mathrm{c\pm}(B)$ data.
  The optimum field $B_\mathrm{opt}\approx -5.38\units{mT}$.
}
\label{fig:IV+Ic(H)}
\end{figure}

The electric transport measurements are done in a 4-point configuration and were performed in liquid He at $T = 4.2 \units{K}$.
The IVC of the device \#A22 at $B=0$ is shown in Fig.~\ref{fig:IV+Ic(H)}(a).
The IVC is RSJ-like with symmetric critical currents and without hysteresis.
To find the optimal working point of the ratchet, we measure the dependence $I_c(B)$.
The field is applied perpendicular to the sample plane by means of a coil.
The $I_\mathrm{c}(B)$ dependence is shown in Fig.~\ref{fig:IV+Ic(H)}(b).
The optimum working point corresponds to the maximum of $\As$, see definition~\eqref{Eq:As.def}.
Using the data from Fig.~\ref{fig:IV+Ic(H)}(b) we display $\As(B)$ on the same plot to determine the value of $B_\mathrm{opt}$, where $\As(B)$ has its maximum $\approx 7$.
Figure \ref{fig:IV+Ic(H)}(a) also shows the IVC of the ratchet at $B=B_\mathrm{opt}$.
The critical currents are rather asymmetric.

\subsection{Quasistatic deterministic drive}
\label{Sec:Drive.det}

\begin{figure}[!p]
  \begin{center}
    \includegraphics[width=0.92\columnwidth]{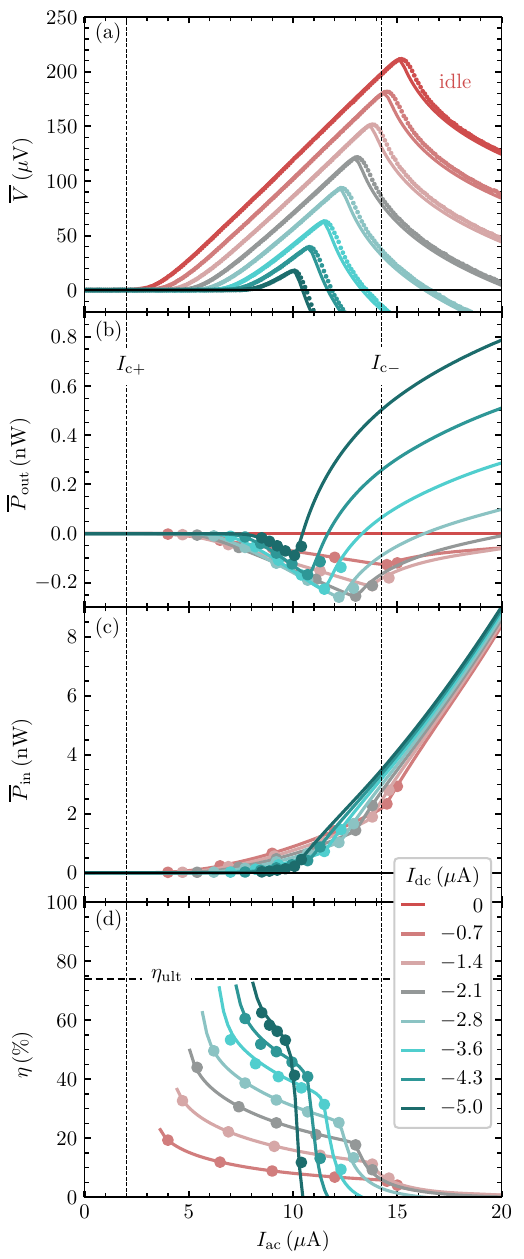}
  \end{center}
  \caption{%
    Performance of JR (sample \#A22) vs. ac drive amplitude $\Iac$ ($f=200\,$Hz) at $B=B_\mathrm{opt}$, for different values of counter current $\Idc\leq 0$.
    (a) rectification curves $\mean{V}(\Iac)$,
    (b) output power $P_\mathrm{out}(\Iac)$,
    (c) input power $P_\mathrm{in}(\Iac)$ and
    (d) efficiency $\eta(\Iac)$.
    Symbols show experimental data. Lines are calculated from the high-resolution experimental IVC as described in the text.
  }
  \label{Fig:RectCurves}
\end{figure}

For demonstration of the ratchet operation in a quasistatic deterministic regime we apply $B=B_\mathrm{opt}$ to induce maximum $\As$ and drive the ratchet with a sinusoidal current
\begin{equation}
  I(t) = \Iac\sin(2\pi f t)
  ,
  \label{Eq:acdrive}
\end{equation}
where we typically use $f=200\units{Hz}$,
i.e., we operate in the adiabatic regime $f\ll f_c$ \cite{Bartussek94,Sterck02,Sterck05}, with the characteristic frequency $f_c=I_cR/\Phi_0\sim300\units{GHz}$ ($R\approx50\units{\Omega}$ is the JJ normal resistance). The $I(t)$ waveform is generated by a programmable DAC card with an update rate of $10\units{kHz}$ (samples/sec).
For a given continuously applied waveform with amplitude $\Iac$ we measure the voltage $\mean{V}$ averaged over one period of the drive $T=1/f=5\units{ms}$, \ie,
\begin{equation}
  \mean{V} = \frac1T \int_0^T V(t)\,d t.
  \label{Eq:meanV}
\end{equation}
Technically this is done by collecting $500$ voltage samples with a sampling rate of $100\units{kHz}$ (interval between the samples $10\units{\mu s}$) by the programmable ADC card.
By repeating the measurement of \mean{V} for different values of $\Iac$ we obtain the \emph{rectification curve} $\mean{V}(\Iac)$ shown in Fig.~\ref{Fig:RectCurves}(a) as the red curve, labeled with ``idle'', which means that the drive is a pure ac drive without any dc counter current, \ie, $\Idc=0$.
One can see that for very low amplitudes $\Iac \lesssim I_\mathrm{c+}$ the rectification is absent as the ac bias point never reaches the voltage branches of the IVC during ac-driving.
For $I_\mathrm{c+} \lesssim \Iac \lesssim |I_\mathrm{c-}|$ the bias point reaches only the positive voltage branch of the IVCs during the positive semi-period, which results in a finite $\mean{V}$ that grows with $\Iac$, until $\mean{V}(\Iac)$ reaches its maximum value $\mean{V}_\mathrm{max}$ at $I_\mathrm{ac,max}\approx|I_\mathrm{c-}|$.
Finally, at $\Iac \approx I_\mathrm{c-}$ the bias point also reaches the negative voltage branch of the IVC during the negative semi-period and the average voltage drops with further increasing $\Iac$ and asymptotically approaches zero for $\Iac\gg|I_\mathrm{c-}|$.
\footnote{
  According to the RSJ-like model\cite{Knufinke12} the linear rectification branch should start at $I_{c+}$ and the maximum of the $\mean{V}(\Iac)$ curve should appear exactly at $I_{c+}$. However, in Fig.~\ref{Fig:RectCurves}(a) the $\mean{V}(\Iac)$ curves seem to be shifted somewhat to the right relative to $I_{c+}$ and $I_{c-}$. This is related to the rounding of our IVC near $I_{c+}$ and $I_{c-}$, which is not captured by the model\cite{Knufinke12}. Namely, the value of the current $I_{m+}$, where the differential resistance reaches its maximum, is somewhat larger than $I_{c+}$. Similarly, $|I_{m-}| \gtrsim |I_{c-}|$. Thus, at $\Iac=I_{c+}$ we do have the onset of rectification, however the voltage becomes substantial (linear branch starts) at $\Iac>I_{m+}$. Similarly, at $\Iac=|I_{c-}|$ the rectification curve starts bending down, but its maximum is reached at $|I_{m+}|$. In the RSJ-like IVC of the model\cite{Knufinke12} $I_{m\pm}=I_{c\pm}$.
}

The rectification is \emph{efficient} \emph{roughly} for \Iac between $I_\mathrm{c+}$ and $I_\mathrm{c-}$.
\footnote{%
  Formally, rectification takes place in an infinite range of $\Iac=I_{c+}\ldots\infty$ where $\mean{V}>0$. However, for $\Iac>|I_{c-}|$ the ratchet works in the so-called ``Sisyphus regime'', where the particle moves in the asymmetric potential back and forth, dissipating a lot with a little net progress. Instead, for $I_{c+}<\Iac<|I_{c-}|$ the particle moves only in the easy (positive) direction and is blocked in the difficult (difficult) direction.
}
If one wants to construct a ratchet which rectifies a large range of input amplitudes, one should ideally have $I_\mathrm{c+} \ll |I_\mathrm{c-}|$ (or $|I_\mathrm{c-}| \ll I_\mathrm{c+}$), \ie, large $\As$.
We observe a maximum rectified voltage $\mean{V}\approx 212\units{\mu V}$, which is one of the best among similar devices, see Tab.~\ref{Tab:cmp}.

Up to now our ratchet is idle, \ie, it does not produce any useful work (output power).
In terms of a particle in an asymmetric periodic potential, this means that the particle is driven by a pure ac drive to the right (easy direction), but stays roughly at the same energy/height.
To produce work one has to load the ratchet.
One possibility \cite{Knufinke12} is to tilt the potential in such a way that the ratchet effect will transport the particle uphill.
In this case one can also address the question ``how strong is the ratchet'', \ie, against which counter tilt the ratchet can still transport the particle.
Experimentally, it is rather easy to tilt the potential just by applying an additional dc bias counter current to the JJ.
If our rectified voltage $\mean{V}>0$ then one needs a counter current $\Idc<0$.
Then, the total applied current is
\begin{equation}
I(t) = \Iac\sin(2\pi f t) + \Idc
.
\label{Eq:I(t).acdc}
\end{equation}

The rectification curves
\begin{equation}
  \mean{V}(\Iac) = \frac1T \int_0^T V(I(t))\,\dd t
  .
  \label{Eq:meanV.sim}
\end{equation}
for several $\Idc$ values are shown in Fig.~\ref{Fig:RectCurves}(a).
One can see that transport against the counter current still takes place i.e., $\mean{V}>0$, although the average voltage (particle speed) and accordingly, the maximum average voltage $\mean{V}_\mathrm{max}$, decreases with increasing $\Idc$.
The rectification range shrinks with increasing $|\Idc|$, and for large $\Iac$ the transport reverses ($\mean{V}<0$).
The stopping current, \ie, the (minimum) counter current, for which the ratchet does not transport anymore in the easy direction for any $\Iac$, is theoretically given by \cite{Knufinke12} $I_\mathrm{stop}=(I_\mathrm{c+} - |I_\mathrm{c-}|)/2\approx -6.1\units{\mu A}$, which agrees quite well with the data presented in Fig.~\ref{Fig:RectCurves}(a).
Note that the ratchet loaded with $\Idc<0$ produces negative $\mean{V}$ at large $\Iac$. This motion in the difficult direction is simply cased by the applied $\Idc$, which overweights the ratchet effect.

Since the ratchet with $\Idc<0$ transports the particle uphill, it produces an (average) output power
\begin{equation}
  \mean{P}_\mathrm{out} =  \frac1T \int_0^T V(I(t))\cdot \Idc\,\dd t = \mean{V}\cdot \Idc .
  \label{Eq:Pout}
\end{equation}
Thus, to obtain plots $\mean{P}_\mathrm{out}(\Iac)$ one should just multiply each $\mean{V}(\Iac)$ curve from Fig.~\ref{Fig:RectCurves}(a) by the corresponding value of $\Idc$.
The result is shown in Fig.~\ref{Fig:RectCurves}(b).
Note that since $\Idc \leq 0$, we formally get $\mean{P}_\mathrm{out}\leq 0$ (if $\mean{V}>0$), which means that we generate power rather than consume it.
Obviously, the idle ratchet ($\Idc=0$) produces $\mean{P}_\mathrm{out}(\Iac) \equiv 0$.
Then by increasing the counter current $|\Idc|$ the amplitude of $\mean{P}_\mathrm{out}(\Iac)$ curves first grows and then decreases in accordance with the product in Eq.~\eqref{Eq:Pout}, where the amplitude of $\mean{V}(\Iac)$ decreases with $|\Idc|$ down to zero at $I_\mathrm{stop}$, at which $ \mean{V}(\Iac)$ and $ P_\mathrm{out}(\Iac)$ change sign.
Note that in the parts of the curves, where $P_\mathrm{out}>0$ (corresponding to parts with $\mean{V}<0$ in Fig.~\ref{Fig:RectCurves}(a)), the ratchet consumes the power from dc counter current source.
Thus, the maximum output power in the regime $P_\mathrm{out}<0$ is expected for an intermediate load between 0 and $I_\mathrm{stop}$.

Similarly, the input power is given by
\begin{equation}
\mean{P}_\mathrm{in} = \frac1T \int_0^T V(I(t))\cdot \Iac\sin(2\pi f t)\,\dd t
.
\label{Eq:Pin}
\end{equation}
Equation \eqref{Eq:Pin} cannot be simplified in a similar way as Eq.~\eqref{Eq:Pout}.
Therefore, to determine $\mean{P}_\mathrm{in}$, one needs simultaneously measured $\Iac\sin(2\pi f t)$ and $V(t)$ profiles.
Those were measured for six different values of $\Iac$ for each $\Idc$ value.
The results are presented in Fig.~\ref{Fig:RectCurves}(c) by symbols.
Roughly, one notes three regions on the plots.
For low $\Iac$ the power vanishes as the ratchet never enters the voltage state.
For intermediate values of $\Iac$ (branches with slight slope in Fig.~\ref{Fig:RectCurves}(c)) the ratchet dissipates only at the positive voltage branch during some part of the positive semi-period of the drive and, finally, for large $\Iac$ (branches with strong slope in Fig.~\ref{Fig:RectCurves}(c)) the ratchet dissipates even more during both positive and negative semi-periods.

In the quasi-static regime of operation, all information on the ratchet performance and its figures of merit are contained in the IVCs. Therefore, having a (high resolution) experimental $V(I)$ (as a list of numerical $I$ and $V$ values), one can calculate all characteristics like $\mean{V}(\Iac)$, $\mean{P}_\mathrm{in}$, $\mean{P}_\mathrm{out}$, \etc for different values of the load $\Idc$ numerically, by ``applying'' the current $I(t)$ given by Eq.~\eqref{Eq:I(t).acdc} to the IVC and calculating the integrals \eqref{Eq:meanV.sim}--\eqref{Eq:Pin} numerically. In this way one can produce quite many points  per curve (esp. relevant for $\mean{P}_\mathrm{in}$). The results are also presented in Fig.~\ref{Fig:RectCurves} as solid lines and show only a minor difference with those measured experimentally.

Finally, having $\mean{P}_\mathrm{out}$ and $\mean{P}_\mathrm{in}$ one can calculate the \emph{thermodynamic efficiency}
\begin{equation}
\eta(\Iac) \equiv - \frac{\mean{P}_\mathrm{out}(\Iac)}{\mean{P}_\mathrm{in}(\Iac)}
.
\label{Eq:eff}
\end{equation}
Note that \emph{thermodynamic efficiency} $\eta$ should not be confused with the ``efficiency'' given by
\begin{equation}
\eta_\mathrm{ult}
= \left|\frac{|I_\mathrm{c-}| - |I_\mathrm{c+}|}{|I_\mathrm{c-}| + |I_\mathrm{c+}|}\right|
\equiv \frac{\As-1}{\As+1}
,
\label{Eq:eff-ult}
\end{equation}
used in many publications.
On the one hand, $\eta_\mathrm{ult}$ just characterizes the degree of asymmetry of critical currents and can be expressed via $\As$ used in this work.
On the other hand, $\eta_\mathrm{ult}$ represents the maximum possible (ultimate \cite{Knufinke12,Goldobin:2016:JRatchet.ModelIVC}) thermodynamic efficiency that can be reached for a given asymmetry theoretically.

A set of $\eta(\Iac)$ curves for different load values are shown in Fig.~\ref{Fig:RectCurves}(d).
Each $\eta(\Iac)$ curve has a sharp maximum just in the beginning of the rectification window, as predicted by the model \cite{Knufinke12,Goldobin:2016:JRatchet.ModelIVC}.
As a function of $\Idc$ the  $\eta(\Iac)$ curves reach their maximum amplitude for large counter current $|\Idc|$ close to $I_\mathrm{stop}$, where the rectification window is tiny.
In this regime the value of $\eta$ approaches its theoretical value $\eta_\mathrm{ult}$ ($\approx 75\%$ for \#A22).

We note that the efficiency is cut (not calculated) for $\Iac<I_\mathrm{c+}$.
In this range, both $\mean{P}_\mathrm{out}\to0$ and $\mean{P}_\mathrm{in}\to0$ (theoretically) so that $\eta$ has a very large uncertainty.
In fact, any measurement (fluctuation) or numerical error in $\mean{P}_\mathrm{in}$ will result in a huge fluctuation of $\eta$.
In other words, $\eta$ will have error bars much larger than the value of $\eta$ itself.
Therefore, the calculation of $\eta$ was not performed, if $\mean{P}_\mathrm{in}$ was smaller than a certain limit (typically $5\units{pW}$).
This is a common problem for the evaluation of the performance of any ratchet operated with small drive amplitudes.

\begin{figure}[!htb]
  \begin{center}
    \includegraphics[width=0.92\columnwidth]{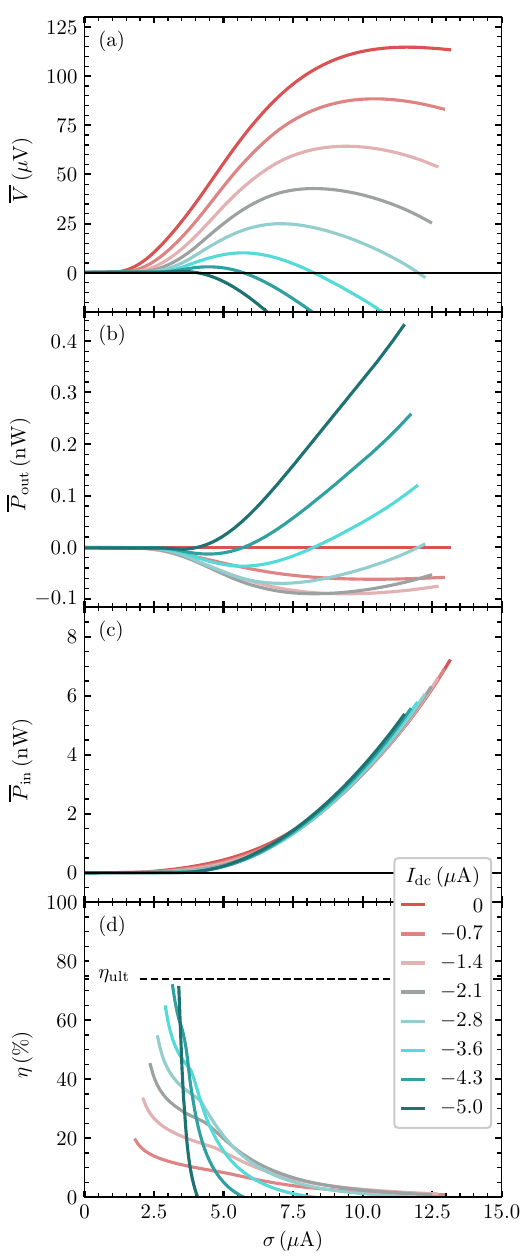}
  \end{center}
  \caption{%
    Performance of JR (calculated from the IVC at $B=B_\mathrm{opt}$ of sample \#A22) vs distribution width (noise amplitude) $\sigma$ of a quasistatic stochastic drive current, for different values of counter current $\Idc\leq 0$.
    (a) Rectification curves $\mean{V}(\sigma)$, (b) $\mean{P}_\mathrm{in}(\sigma)$ and (c) $\mean{P}_\mathrm{out}(\sigma)$ and (d) $\eta(\sigma)$.
    For the calculations we used a $3\sigma$ cutoff during integration, see Eq.~\eqref{Eq:G:Vmean_intI} and the following text.
  }
  \label{Fig:Stoch:RectCurves}
\end{figure}

\subsection{Quasistatic stochastic drive}
\label{Sec:Drive.stoch}

In this section, starting from the experimentally measured asymmetric IVC, we numerically calculate rectification of a random driving force $I(t)=\Xi(t)$ with a Gaussian distribution of the probability density
\begin{equation}
G(I,\sigma) = \frac{1}{\sqrt{2\pi}\sigma}\exp\left(-\frac{I^2}{2\sigma^2}\right)
.
\label{Eq:Gauss}
\end{equation}
Here the distribution width $\sigma$ plays a role of the amplitude of the noise.
In fact, if one has a realization of a random process $\Xi(t)$ with amplitude $1$ (dimensionless), the process $\sigma\cdot\Xi(t)$ has the distribution width $\sigma$.
We will consider \emph{quasistatic} noise, \ie, each random current value $\sigma\cdot\Xi(t_i)$ is applied and kept long enough to measure the voltage $V(\sigma\cdot\Xi(t_i))$ (IVC), then the next random current value $\sigma\cdot\Xi(t_{i+1})$ is applied and so on. In other words, the bandwidth of this noise is much smaller than the typical Josephson frequencies (so that the concept of IVC makes sense) and our measurement setup bandwidth.
Then the average voltage $V_\mathrm{dc}$ (to distinguish it from the voltage $\mean{V}$ averaged over one period in the case of a deterministic periodic drive) is calculated.
Due to quasi-staticity, the result can be again obtained just from the experimentally measured IVC numerically as
\begin{eqnarray}
  V_\mathrm{dc}(\sigma)
  &=& \lim_{T\to\infty}\frac{1}{T}\int_0^T V(\sigma\cdot\Xi(t)+\Idc)\,\mathrm{d}t
  \nonumber\\
  &=& \lim_{N\to\infty}\frac{1}{N}\sum_{i=1}^N V(\sigma\cdot\Xi(t_i)+\Idc).
  \label{Eq:G:Vmean_intt}
\end{eqnarray}
Numerically, the expression \eqref{Eq:G:Vmean_intt} needs a long integration time to converge.
However, for ergodic random processes this can be drastically simplified to a convolution of the Gaussian random distribution with the IVC:
\begin{equation}
  V_\mathrm{dc}(\sigma) = \int_{-\infty}^{+\infty} G(I,\sigma)V(I+\Idc) \,\mathrm{d}I.
  \label{Eq:G:Vmean_intI}
\end{equation}
Note that the experimental $V(I)$ is available (measured) only in certain limited current range.
Therefore, to perform the integration in Eq.~\eqref{Eq:G:Vmean_intI} in practice we integrate from $-3\sigma$ to $+3\sigma$, thus, cutting off the tails of the Gaussian distribution.

A set of rectification curves $V_\mathrm{dc}(\sigma)$ for different counter currents $\Idc$ are shown in Fig.~\ref{Fig:Stoch:RectCurves}(a).
Qualitatively the curves look very similar to the $\mean{V}(\Iac)$ curves for a deterministic sinusoidal ac drive.
However, the maximum rectified voltage (in the idle regime) became almost twice smaller.
The curves are also much more smooth, which is understandable considering the Gaussian distribution of the noise.
However, the stopping force does not change, which is easy to understand from the IVC.
Namely, the applied counter current $\Idc$ basically shifts the origin of the IVC so that the IVC becomes more symmetric.
Roughly the rectification vanishes when the positive and negative critical currents become equal, \ie, $I_\mathrm{c+}-\Idc \approx |-I_\mathrm{c-}+\Idc|$ regardless what kind of drive is applied.

Furthermore, a set of $\mean{P}_\mathrm{in}(\sigma)$, $\mean{P}_\mathrm{out}(\sigma)$ and $\eta(\sigma)$ curves for different values of $\Idc$ are shown in Fig.~\ref{Fig:Stoch:RectCurves}(b)--(d).
One can see that here the curves are also similar to the ones with a sinusoidal drive.
In particular, the efficiency $\eta(\sigma)$ for strong load (counter current) still tends to almost reach the ultimate efficiency $\eta_\mathrm{ult}$.

We would like to point out that rectification of the quasistatic Gaussian noise does not contradict the second law of thermodynamics. In fact, equilibrium thermal fluctuations produce a noise current with a Gaussian probability density and amplitude
\begin{equation}
  \sigma = \sqrt{\frac{2k_BT \cdot \Delta f}{R}},
  \label{Eq:sigma.therm}
\end{equation}
where $\Delta f$ is the bandwidth of the system (width of the white noise spectrum).
Quasistatic noise essentially means that $\Delta f\ll f_\mathrm{setup}\ll f_\mathrm{J}$. For example, for $\Delta f=10\units{kHz}$ according to Eq.~\eqref{Eq:sigma.therm} the amplitude of the thermal noise is $\sigma \sim 0.15\units{nA}$, while we apply the amplitudes at least $10^4$ times larger, see Fig.~\ref{Fig:Stoch:RectCurves}. This essentially means that our noise is not thermal.

\Sec{Conclusions}
\label{Sec:Conclusions}

We demonstrated the design, fabrication and quasistatic operation of a Josephson diode ``drawn'' into a YBCO thin-film micro-bridge using He-FIB.
The ratchet shows record figures of merit, see Tab.~\ref{Tab:cmp}, and its direction of rectification depends on the sign of the applied (optimum) field.
In particular, the ratchet occupies an \emph{essential} area $\approx 1\units{\mum^2}$, which is the smallest in Tab.~\ref{Tab:cmp}.
At the optimum magnetic field the ratchet shows an impressive asymmetry $\As\approx 7$, close to the similar design based on Nb\cite{Golod:2019:SFS-JJ.SPS,Golod:2022:JosDiode@B=0}.
As a consequence, it demonstrates a maximum rectified voltage $\mean{V}\approx212\units{\mu V}$ for a sine-drive and a  $\mean{V}\approx115\units{\mu V}$ (calculated from experimental IVC) for a random Gaussian drive.
In both cases the ratchet shows a large stopping force (in accordance with the value of $\As$, see Refs.~\onlinecite{Knufinke12,Goldobin:2016:JRatchet.ModelIVC}), and the thermodynamic efficiency approaching the theoretical limit (ultimate efficiency) in certain regimes.
However, there is a general trade-off between maximum output power and maximum efficiency that occur at different values of parameters (drive amplitude and counter force).
The only ratchet in Tab.~\ref{Tab:cmp} that shows larger output power (estimated) is the one reported in Ref.~\onlinecite{Knufinke12}. There, the ratchet design was a large ALJJ with very high $I_c$.

Preliminary measurements show that the ratchets discussed in the present paper operate at temperatures up to $\sim40\units{K}$, where the critical currents of JJ tend to zero, while the thermal energy increases by one order of magnitude. Detailed results for the noise-driven ratchet will be published elsewhere.

\acknowledgments

This work was funded by the Deutsche Forschungsgemeischaft (DFG) via projects No.~GO-1106/6 and KL-930/17.
A.~J.~thanks his father for financial support during his study.
We thank M.~Turad and R.~L\"offler for invaluable help with ``Orion Nanofab''.


\appendix

\Sec{Estimating $\lambda_J$}
\label{Sec:EstLambda_J}

To estimate $\lambda_J$ we use the usual expression \cite{Barone82}
\begin{equation}
\lambda_J = \sqrt{\frac{\Phi_0}{2\pi\cdot \ell_J \cdot j_c}}
,
\label{Eq:lambda_J.def}
\end{equation}
rewritten explicitly isolating $\ell_J$ --- an inductance per JJ length times thickness $d$ of the superconducting electrodes forming the JJ when the current flows along the Josephson barrier (units are $\units{H=V \cdot s/A}$, like for inductance).
\footnote{In a tri-layer (strip-line) geometry, $\ell_J$ is known as  ``inductance per square'' of superconducting (thin film) electrodes  forming the JJ.}
Thus, in our planar case, the total inductance of the two pieces of superconducting films on both sides of the barrier is given by $L_\Sigma=\ell_J\cdot L/d$.

In our particular case $\ell_J=2\ell$, where $\ell$ corresponds to one superconducting electrode.
The JJ barrier thickness (created by He-FIB) is considered to be negligible in comparison with the electrode width $w$.
The total inductance $L_\mathrm{el}$ of a piece of superconducting electrode of length $L$ and (film) thickness $d$ is given by $L_\mathrm{el}=\ell\cdot L/d$, \ie, $L_\Sigma = 2L_\mathrm{el}$


In our ultra-thin-film limit ($d=30\units{nm}$, while the London penetration depth $\lambda_\mathrm{L}\approx250\units{nm}$) we can safely assume that the inductance is purely kinetic.
So when the current flows in the superconducting film along the barrier of the JJ, the associated kinetic energy is given by
\begin{equation}
E_\mathrm{k} = n_s\frac{mv_s^2}{2}\cdot V'
,
\label{Eq:Ek.a}
\end{equation}
where $V'=Lwd$ is the volume of the film, while $m$, $n_s$ and $v_s$ are mass, concentration and average velocity of ``superconducting'' electrons, respectively.

Using the relation for the supercurrent density $j_s=n_s v_s e$, we can rewrite Eq.~\eqref{Eq:Ek.a} as
\begin{equation}
  E_\mathrm{k}
  = \frac{mj_s^2}{2n_s e^2}\cdot Lwd
  = \frac{m I_s^2}{2n_s e^2 w^2 d^2}\cdot Lwd
  = \frac{m L}{n_s e^2 w d} \frac{I_s^2}{2},
  \label{Eq:Ek.b}
\end{equation}
where we have introduced the supercurrent $I_s = j_s \cdot w \cdot d$, which assumes a homogenous $j_s$ in our electrodes with $w\ll \lambda_\mathrm{eff}=\lambda_\mathrm{L}^2/d$.
We remind that in the framework of the London theory
\begin{equation}
\lambda_\mathrm{L}^2 = \frac{m}{\mu_0 n_s e^2}
.
\label{Eq:lambda.def}
\end{equation}
Thus,
%
\begin{equation}
  E_\mathrm{k} = \mu_0\lambda_\mathrm{L}^2 \frac{ L}{w d} \frac{I_s^2}{2}.
  \label{Eq:Ek.lambda}
\end{equation}
The kinetic inductance is therefore
\begin{equation}
  L_k = \mu_0\lambda_\mathrm{L}^2 \frac{ L}{w d},
  \label{Eq:Lk}
\end{equation}
\ie,
\begin{equation}
\ell = \mu_0\lambda_\mathrm{L}^2 \frac{1}{w}
,
\label{Eq:ell.one}
\end{equation}
For our parameter $w=200\units{nm}$ we get $\ell_J=2\ell = 0.79\units{pH}$. This, according to Eq.~\eqref{Eq:lambda_J.def} with $j_c=27\units{kA/cm^2}$, gives $\lambda_J\approx 1.2\units{\mu m}$.

Quantitatively, the inductance of the JJ barrier of length $L$
(Josephson inductance at infinitesimal current, see Ref.~\onlinecite{Barone82})
is given by
\begin{equation}
  L_J = \frac{\Phi_0}{2\pi I_c} = \frac{\Phi_0}{2\pi j_c \cdot L \cdot d}.
  \label{Eq:L_J}
\end{equation}
Assuming that $L\to\infty$ ($L\gg\lambda_J$) $\lambda_J$ is the characteristic length, across which the bias current injected from the edge distributes and tunnels through the JJ barrier. At this length the Josephson inductance of the $\lambda_J$-piece of the barrier is equal to the inductance of the $\lambda_J$-piece of both electrodes, \ie,
\[
  \left.L_J\right|_{L=\lambda_J} = 2 \left.L_k\right|_{L=\lambda_J}.
\]
Inserting, $L_k$ and $L_J$ from Eqs.~\eqref{Eq:Lk} and \eqref{Eq:L_J}, we obtain
\[
  \frac{\Phi_0}{2\pi j_c \cdot \lambda_J \cdot d} =  2\mu_0 \lambda_\mathrm{L}^2\frac{ \lambda_J}{w d}
\]
From here
\[
  \lambda_J^2  = \frac{\Phi_0 w d}{2\pi j_c d 2\mu_0\lambda_\mathrm{L}^2}
  = \frac{\Phi_0 w}{2\pi j_c 2\mu_0\lambda_\mathrm{L}^2}
  = \frac{\Phi_0  }{2\pi j_c \ell_J},
\]
\ie, exactly as given by Eq.~\eqref{Eq:lambda_J.def} with $\ell_J=2\ell$ from Eq.~\eqref{Eq:ell.one}.


\Sec{In-line geometry.}
\label{Sec:InlineNum}

\subsection{Derivation of $I_c(B)$}

\begin{figure}[!htb]
  \begin{center}
    \includegraphics*{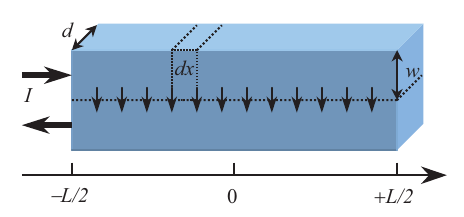}
  \end{center}
  \caption{The JJ of in-line geometry consisting of two superconducting films and a very thin barrier (dotted). The arrows indicate the current $I$ injected/collected at the left edge and flowing through the barrier.}
  \label{Fig:inlineJJ}
\end{figure}

Following Ref.~\onlinecite{Barone82}, we consider a JJ of inline geometry spanning along $x$ from $-L/2\ldots +L/2$. The bias current $I$ is injected to and collected from the left side of the superconducting electrodes. We assume that the JJ is short, \ie, $L \lesssim 4\lambda_J$, and, for now, we assume zero applied magnetic field. The bias current distributes along the whole JJ length $L$ to tunnel through the barrier. Since the JJ is short, we assume that the Josephson phase $\phi(x)$ is almost constant across the barrier. In this case, the Josephson current density across the barrier $j_s(x)$ is constant too. Then, we write a current continuity (1st Kirchoff) equation at an arbitrary point $x$ inside the JJ.
\[
  I_L(x+\dd x)-I_L(x)=j_s(x)\cdot \dd x\cdot d,
\]
which, for $\dd x\to0$ gives
\begin{equation}
  \fracp{I_L(x)}{x} \equiv I'_L(x) = j_s(x) \cdot d,
  \label{Eq:I_L.cont}
\end{equation}
where $I_L(x)$ is the current flowing along the top superconducting electrode. Considering our inline biasing scheme the boundary conditions (BCs) for $I_L(x)$ are
\begin{equation}
  I_L(-L/2) = I,\quad I_L(+L/2) = 0.
  \label{Eq:I_L.BCs}
\end{equation}

By solving Eq.~\eqref{Eq:I_L.cont} with BCs \eqref{Eq:I_L.BCs} we get an explicit expression for
\begin{equation}
  I_L(x) = -j_s d \sbraces{x-L/2} = -I \sbraces{\frac{x}{L}-\frac12}
  , \label{Eq:I_L(x)}
\end{equation}
where we have used the obvious fact that the whole bias current finally tunnels through the JJ, \ie, $I=j_s d L$.

The current $I_L(x)$ flowing through the kinetic inductance $\dd L_k$ of the $\dd x$ piece of the top/bottom electrode creates a phase difference (of the macroscopic wavefunction of the electrode)
\[
  \theta(x+\dd x)-\theta(x)
  = \frac{2\pi}{\Phi_0}\cdot \dd L_k I_L(x)
  = \frac{2\pi}{\Phi_0}\cdot \mu_0\lambda_\mathrm{L}^2 \frac{\dd x}{wd} I_L(x),
\]
which for $\dd x\to0$ gives
\begin{equation}
  \theta'(x) = \frac{2\pi}{\Phi_0}\cdot \mu_0\lambda_\mathrm{L}^2 \frac{I_L(x)}{wd}
  , \label{Eq:theta'(x)}
\end{equation}
where $\mu_0\lambda_\mathrm{L}^2$ is the specific kinetic inductance, see Appendix~\ref{Sec:EstLambda_J}.

By substituting the expression \eqref{Eq:I_L(x)} into Eq.~\eqref{Eq:theta'(x)} and solving it, we obtain
\begin{equation}
  \theta(x) = \frac{2\pi}{\Phi_0}\cdot \frac{\mu_0\lambda_\mathrm{L}^2}{wd} I\braces{\frac{x}{2}-\frac{x^2}{2L}}
  . \label{Eq:theta(x)}
\end{equation}
For the moment we omitted the integration constant as it will be added later when we consider and maximize the supercurrent.

The Josephson phase $\phi(x)$ is the difference of the phases $\theta_1(x)$ and $\theta_2(x)$ in electrodes 1 and 2. Since the electrode currents $I_{L1}(x)$ and $I_{L2}(x)$ flow in opposite directions, with accuracy of a constant one can write $\theta_2(x)=-\theta_1(x)$. Therefore, $\phi(x)=2\theta_1(x)$, see Eq.~\eqref{Eq:theta(x)}. Following Ref.~\onlinecite{Barone82}, we ignore the parabolic bending of the Josephson phase and keep only the linear term (global behavior), \ie,
\begin{equation}
  \phi(x) \approx \frac{2\pi}{\Phi_0}\cdot \frac{\mu_0\lambda_\mathrm{L}^2}{wd} \cdot x \cdot I
  . \label{Eq:phi(x).self.lin}
\end{equation}
This dependence is very similar to the linear Josephson phase created by an applied magnetic field $B$ perpendicular to the film plane
\begin{equation}
  \phi'(x) = \frac{2\pi}{\Phi_0}d_\mathrm{eff}B,
  \text{ or }
  \phi(x) = \frac{2\pi}{\Phi_0}d_\mathrm{eff}B x = 2\pi f \frac{x}{L}
  , \label{Eq:phi(x).B}
\end{equation}
where we have introduced the normalized flux $f\equiv\Phi/\Phi_0$, where $\Phi\equiv B d_\mathrm{eff} L$ is the total magnetic flux threading the JJ. In our geometry $d_\mathrm{eff}=2\lambda_\mathrm{L}\tanh{w/\lambda_\mathrm{L}}\approx 2w$. The linear Josephson phase in Eq.~\eqref{Eq:phi(x).self.lin} is proportional to the bias current $I$. In Ref.~\onlinecite{Barone82} this is called a self-field effect. In our case the effect is exactly the same, but due to kinetic nature of the electrode's inductance the magnetic field induced by the bias current $I$ as such is not present.

To obtain the total supercurrent, but now also including an applied magnetic field, we have to add the phase gradients resulting from both bias current and applied field. The ansatz reads
\begin{equation}
  \phi(x) = 2\pi(f + f_I \cdot i) \frac{x}{L} + \phi_0
  , \label{Eq:phi_tot.norm}
\end{equation}
where $i=I/I_{c0}$ is the normalized bias current and
\begin{equation}
  f_I \equiv \frac{\mu_0\lambda_\mathrm{L}^2 \frac{L}{wd}I_{c0}}{\Phi_0} = \frac{L_K I_{c0}}{\Phi_0} = \frac{\Phi_{I_{c0}}}{\Phi_0}
  , \label{Eq:}
\end{equation}
characterizes the strength of the ``self-field'' effect (Josephson-phase gradient due to bias current). It is defined as pseudo-flux $\Phi_{I_{c0}}$ through the kinetic inductance $L_k = \mu_0\lambda_\mathrm{L}^2 \frac{L}{wd}$ of the whole top electrode when one sends a current $I_{c0}$ through it.

Following the standard procedure to find the total supercurrent
\[
  I_s(f,i,\phi_0) = d \cdot \int_{-L/2}^{+L/2} j_c\sin\phi(x)\,\dd x
\]
with $\phi(x)$ given by Eq.~\eqref{Eq:phi_tot.norm} and then maximizing $I_s(f,i,\phi_0)$ with respect to $\phi_0$ we get for the normalized critical current $i_c=I_c/I_{c0}$
\begin{equation}
  i_c(f,i_c) = \pm\frac{\sin\sbraces{\pi (f+f_I i_c)}}{\pi (f+f_I i_c)}
  , \label{Eq:i_c(f).imp}
\end{equation}
where $I_{c0}\equiv j_c L d$ is the maximum possible critcal current through the barrier at $f=f_I=0$.
Note, that by using the definition \eqref{Eq:lambda_J.def} of $\lambda_J$, one can rewrite the expression for $f_I$ in a very simple and understandable form, namely
\begin{equation}
  f_I = \frac1{4\pi}\frac{L^2}{\lambda_J^2}.
  \label{Eq:f_I(L,lambda_J)}
\end{equation}

\subsection{Optimal parameters}

\begin{figure}[htb]
  \begin{center}
    \includegraphics{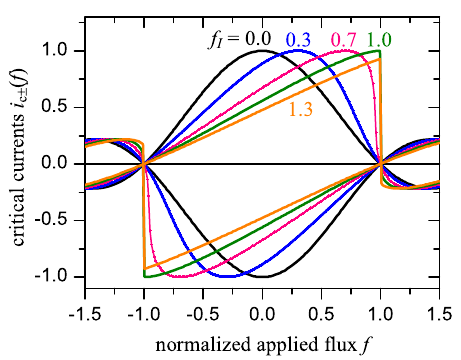}
  \end{center}
  \caption{A family of asymmetric $i_\mathrm{c\pm}(f)$ curves for different $f_I$ obtained by solving the eq.~\eqref{Eq:i_c(f).imp} numerically.}
  \label{Fig:i_c(f)@f_I}
\end{figure}

The final $i_c(f)$ dependence is given by the implicit expression \eqref{Eq:i_c(f).imp}. For fixed values of $f_I$ and $f$ we solved this equation numerically to obtain $i_\mathrm{c\pm}(f)$ plots for several different values of $f_I$, see Fig.~\ref{Fig:i_c(f)@f_I}. With increasing $f_I$ the $i_\mathrm{c\pm}(f)$ plots depart from a symmetric Fraunhofer pattern ($f_I=0$ curve) and become skewed, however they are still point-symmetric with respect to the origin. As $f_I$ grows the two critical currents $i_\mathrm{c+}(f)$ and $i_\mathrm{c-}(f)$ become rather different for $|f|$ somewhat below 1 thus giving high asymmetry. At $f_I > f_I^*=\pi/4\approx0.785$ one of the $i_\mathrm{c\pm}(f)$ curves develops a discontinuous jump at $f=\pm1$ from high $i_c$ (absolute) values to low ones.

\begin{figure}[htb]
  \begin{center}
    \includegraphics{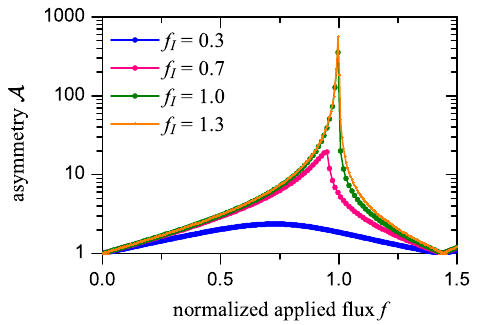}
  \end{center}
  \caption{Asymmetry parameter $\As(f)$ for different values of $f_I$ obtained from the curves in Fig.~\ref{Fig:i_c(f)@f_I} using the definition \eqref{Eq:As.def}.}
  \label{Fig:Asym(f)}
\end{figure}

\begin{figure}[htb]
  \begin{center}
    \includegraphics{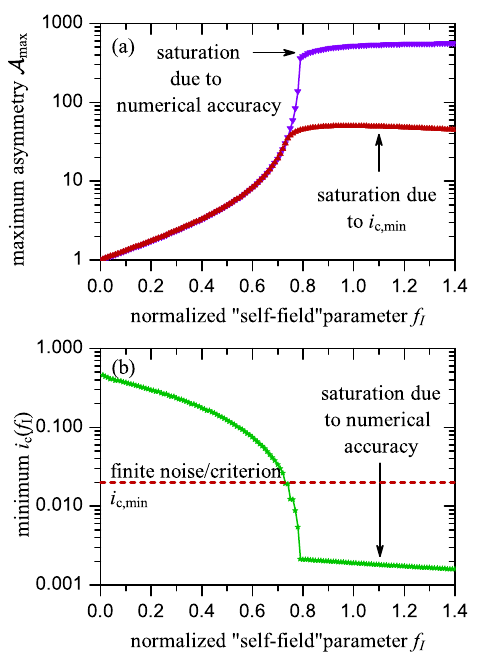}
  \end{center}
  \caption{%
    (a) Maximum asymmetry parameter $\As_\mathrm{max}(f_I)$ taken at $f_\mathrm{opt}$ for each $f_I$. Violett curve is calculated according to theoretical model. Bordeaux curve is calculated taking into account that $i_\mathrm{c,min}$ cannot become smaller than 0.02.
    (b) The value if the minimum (by absolute value) critical current $i_\mathrm{c,min}$ at the optimal point $f_\mathrm{opt}$ as a function of $f_I$. Dashed line shows (as an example) the experimental limitation on $i_\mathrm{c,min}$, which then used to calculate the bordeaux curve in (a).
  }
  \label{Fig:AsymMax}
\end{figure}

To be more specific, in Fig.~\ref{Fig:Asym(f)} we have plotted the dependence of the asymmetry $\As(f)$ (only for positive $f$, negative $f$ are similar) for different values of $f_I$. There is an optimum applied field $f_\mathrm{opt}$, for which $\As$ has a maximum $\As_\mathrm{max}\equiv\As(f_\mathrm{opt})$. Moreover, $\As_\mathrm{max}$ and $f_\mathrm{opt}$ increase with increasing $f_I$ monotonically and for $f_I \ge f_I^*$ $\As(f)$ develops a jump at $f=1$ related to the discontinuity of the $i_{c+}(f)$ branch.

Finally, to see, which values of asymmetry can be reached in principle, we show $\As_\mathrm{max}(f_I)$ in Fig.~\ref{Fig:AsymMax}(a). $\As_\mathrm{max}(f_I)$ rapidly increases with increasing $f_I$ and formally diverges at $f_I^*$. The saturation value of $\As_\mathrm{max}\sim600$ for $f_I \ge f_I^*$ is defined solely by numerical accuracy.

Thus, in the framework of the inline JJ model introduced here, for optimum applied flux $f\approx1$ and large enough in-line geometry parameter $f_I \ge f_I^*$, one can obtain almost unlimited asymmetry values.

\subsection{Possible limitations}
\label{Sec:AsymLimits}

The model used above is an idealization.
First, our initial assumption of a short JJ means that $f_I$ defined by Eq.~\eqref{Eq:f_I(L,lambda_J)} cannot be very large. However, we obtain very high $\As$ values already for $f_I=f_I^*$. This corresponds to $L=\pi\lambda_J$, which is still within the  short limit. However, the numerical solution of the sine-Gordon equation to obtain $i_c(f,f_I)$ (not presented) shows that at $L\approx (3 \ldots 5) \lambda_J$ the minimum critical current does not approach zero closely, so that $\As_\mathrm{max}$ just reaches values of about $5 \ldots 6$.
Second, our approximation (linearization) of the Josephson phase in Eq.~\eqref{Eq:phi(x).self.lin} may be a reason of the extremely high $\As$ obtained. The huge values of $\As$ occur at $f_\mathrm{opt}\approx1$ when one of the critical currents (almost) vanishes while the other one stays finite. However, when the bias current is small, the ``self-field'' term (both linear and nonlinear one) plays no role for the branch with vanishing $i_c$. Instead, it may make a certain small correction to the branch with the high $i_c$. Altogether, high values of $\As$ weakly depend on non-linear term in Eq.~\eqref{Eq:theta(x)}.

There are several practical (experimental) limitations that do not allow one to achieve very high values of $\As$ because it is very difficult experimentally to obtain vanishing $i_\mathrm{c}(f)$ (one of the two) at $f_\mathrm{opt}$. In Fig.~\ref{Fig:AsymMax}(b) we plot the smallest critical current $|i_\mathrm{c,min}(f_I)|$ taken at $f_\mathrm{opt}$. For $f_I \ge f_I^*$ the value of $i_\mathrm{c,min}(f_I)\sim0.002$, which again is defined by numerical accuracy. In experiment, due to a number of reasons $i_\mathrm{c,min}$ cannot be so small. For example, (a) non-uniformity in $j_c(x)$, typical for YBCO-based JJs, results in an $i_c(f)$ pattern, where the minima are lifted relative to the $i=0$ level. Another reason (b) is that in experiment $I_c(B)$ is measured with some finite voltage criterion $V_\mathrm{cr}$ (typically $1\ldots2\units{\mu V}$ due to noise and limited resolution of the equipment), which results in a background $I_c$ level $I_c^\mathrm{bg}\approx V_\mathrm{cr}\cdot R_n$, where $R_n$ is the normal resistance of the JJ. If we assume that $i_\mathrm{c,min}(f_I)$ below 0.02 cannot be measured, then the maximum value of asymmetry $\As_\mathrm{max} \approx 1/0.02=50$ at best. In Fig.~\ref{Fig:AsymMax}(b), as an example, we show this level by a dashed horizontal line. If the theoretical value $i_\mathrm{c\pm}(f_I)$ becomes lower, we then use $0.02$ for calculation of the asymmetry $\As_\mathrm{max}$. The result is shown in Fig.~\ref{Fig:AsymMax}(a) (bordeaux curve). $\As_\mathrm{max}$ for $f_I \ge f_I^*$ is substantially reduced from (formally) infinity down to $40..50$.


\bibliography{this,ratch,LJJ,JJ}

\begin{thebibliography}{54}
\bibitem [{Tab({\natexlab{a}})}]{TabNote:Knufi}%
  \BibitemOpen
  \bibinfo {note} {In Ref.~\onlinecite{Knufinke12}, only $\mean{V}$ was
  measured. From Fig.~2 and Fig.~3, we determine that $I_{\mathrm{c_-}} = -0.5$
  and $I_{\mathrm{c_+}} = 0.3$ in units of $I_0 = 2.8 \units{mA}$. We then find
  $I_{\mathrm{dc}}^{\mathrm{opt}} \approx - 0.06$ corresponds to maximum
  $\mean{P}_{\mathrm{out}}$ from eq.~(18), and substitute this value in
  eq.~(17) to obtain maximum $\mean{P}_{\mathrm{out}}$. Note that, the
  quantities in eq. (17) are normalized. Therefore, to obtain the power in
  physical units, we multiply by $V_1\cdot I_0 = 280 \units{nW}$, where $V_1
  \approx 100 \units{\mu V}$.}\BibitemShut {Stop}%
\bibitem [{Tab({\natexlab{c}})}]{TabNote:Golod}%
  \BibitemOpen
  \bibinfo {note} {Quoted asymmetry and other figures of merit are given for
  zero magnetic field. At the optimum magnetic field the asymmetry $\sim
  10$.}\BibitemShut {Stop}%
\bibitem [{Tab({\natexlab{d}})}]{TabNote:RectDrv}%
  \BibitemOpen
  \bibinfo {note} {The authors used rectangular drive. For sine-drive the
  average voltage will be (substantially) lower.}\BibitemShut {Stop}%
\end{thebibliography}

\end{document}